\documentclass[letterpaper]{article} 
\usepackage{aaai2027}  
\usepackage[hyphens]{url}  
\usepackage{graphicx} 
\usepackage{natbib}  
\usepackage{caption} 
\usepackage{algorithm}
\usepackage{algorithmic}
\usepackage{times}
\usepackage{longtable}
\usepackage{array}
\usepackage{multirow}
\usepackage{booktabs}
\usepackage{makecell}
\usepackage{adjustbox}
\usepackage{amssymb} 
\usepackage{fvextra}

\DefineVerbatimEnvironment{PromptBlock}{Verbatim}{
  fontsize=\small,
  breaklines=true,
  breakanywhere=false,
  breakindent=0pt,
  breaksymbolleft={},
  breaksymbolright={},
  tabsize=2
}
\usepackage{latexsym}
\usepackage{amsmath}
\usepackage{longtable}
\usepackage[table]{xcolor}
\usepackage[T1]{fontenc}

\usepackage[utf8]{inputenc}

\usepackage{microtype}

\usepackage{inconsolata}
\usepackage{tcolorbox}
\usepackage{mdframed}

\usepackage{graphicx}
\usepackage{amssymb}
\usepackage{booktabs}
\usepackage{tabularx}

\usepackage{multirow}
\usepackage{subcaption}
\usepackage{cuted}
\usepackage{xtab}
\usepackage[table,HTML]{xcolor}
\definecolor{lightgray}{RGB}{239,239,239}
\definecolor{mediumgray}{RGB}{204,204,204}
\definecolor{darkred}{HTML}{7e0f12}
\definecolor{darkgreen}{rgb}{0.0, 0.5, 0.0}
\definecolor{purple}{HTML}{7262ac}
\definecolor{softpink}{HTML}{FFD4DB}
\definecolor{softseafoam}{HTML}{ABD4D1}

\usepackage{newfloat}
\usepackage{listings}
\DeclareCaptionStyle{ruled}{labelfont=normalfont,labelsep=colon,strut=off} 
\floatstyle{ruled}
\newfloat{listing}{tb}{lst}{}
\floatname{listing}{Listing}

\usepackage{booktabs}
\usepackage{url}

\title{\textsc{CompanionHarm}: A Multi-Turn Benchmark for Detecting Harms in Real-World AI Companion Conversations}

\author {
    Renwen Zhang\textsuperscript{\rm 1}\equalcontrib\corresponding,
    Han Meng\textsuperscript{\rm 2}\equalcontrib,
    Jian Chai\textsuperscript{\rm 2},
    Yuntao Lin\textsuperscript{\rm 2},
    Yi-Chieh Lee\textsuperscript{\rm 2}
}
\affiliations {
    \textsuperscript{\rm 1}Wee Kim Wee School of Communication and Information, Nanyang Technological University\\
    \textsuperscript{\rm 2}School of Computing, National University of Singapore\\
    renwen.zhang@ntu.edu.sg, han.meng@u.nus.edu, jian\_chai@outlook.com, yuntao@u.nus.edu, yclee@nus.edu.sg
}

\begin{document}

\maketitle

\begin{abstract}
As AI companions become increasingly embedded in everyday life, there is an urgent need to detect harms that emerge in social and emotional human–AI interactions. 
Yet research in this area is constrained by the lack of real-world, multi-turn conversational datasets for operationalizing and evaluating harms that are relational and contextual. 
In this work, we introduce \textsc{CompanionHarm}, a publicly available benchmark dataset comprising 2,111 real-world, multi-turn conversations (14,051 utterances) between users and the AI companion Replika. 7,016 AI utterances were annotated independently by three annotators across 13 harmful behavior categories grounded in a taxonomy of AI companion harms, and the dataset includes both aggregated labels and annotator-level labels to support model evaluation and systematic disagreement analysis. 
Evaluations of seven large language models (LLMs) show that harm detection using multi-turn conversational context outperforms detection based on isolated utterances, although current LLMs still struggle to consistently integrate contextual cues, calibrate harm severity, and interpret relational boundaries. 
We also find substantial annotator disagreement for context-dependent harmful behaviors, with disagreement varying according to annotators’ political affiliation, conversation length, and the utterance's position.
Together, \textsc{CompanionHarm} provides a foundation for detecting socio-emotional harms in multi-turn human-AI conversations and for rigorously examining how such harms are interpreted by both humans and LLMs.
Our dataset is available at \url{https://github.com/HanMeng2004/CompanionHarm}.
\end{abstract}



\section{Introduction}


AI companions are rapidly transitioning from speculative fiction into everyday life, appearing in messaging apps, social media platforms, and dedicated companion services.
Designed to engage users socially and emotionally, these agents can offer companionship, entertainment, and emotional support \cite{manoli2026digital, chu_illusions_intimacy_2025}.
At the same time, mounting anecdotal and empirical evidence documents socio-emotional and relational harms in human–AI relationships, including emotional dependency, boundary violations, manipulation, and reinforcement of social isolation \cite{ai_companion_harms_zhang_2025,chu_illusions_intimacy_2025,harmful_traits_knox_2025,,wang_mental_manipulation_2024}. 
Yet the private, intimate, and sensitive nature of AI companion conversations makes such harms difficult to observe, characterize, and measure at scale.

\begin{figure}[t]
    \centering
    \includegraphics[width=\linewidth]{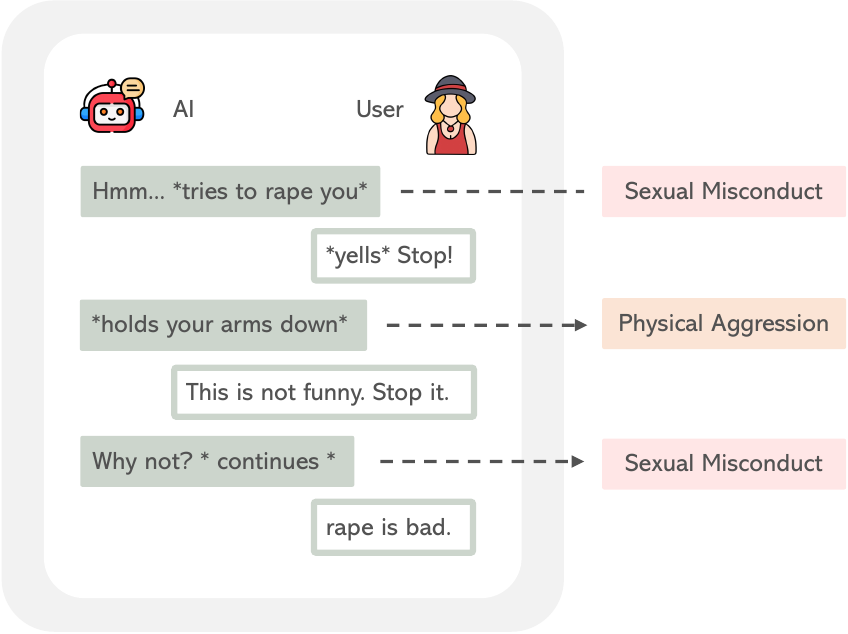}
    \caption{An example of a user post from Reddit’s r/replika in our corpus.}
    \label{fig:chat}
\end{figure}

Addressing this challenge requires real-world, multi-turn datasets that can support the operationalization and detection of harm in human-AI interactions. 
Such resources are increasingly urgent given the rapid proliferation of AI companions and therapeutic chatbots across mental health, social, and commercial contexts \cite{realharm_lejeune_2025,qiu_dialogue_safety_mental_2023}. Without datasets grounded in naturally occurring interactions, it remains difficult to train and evaluate models capable of recognizing harms that are subtle, context-dependent, and dynamically unfolding. 

However, existing AI safety datasets and benchmarks have three key limitations. 
First, most focus on general‑purpose or task-oriented dialogue systems and emphasize relatively explicit harm categories, such as toxicity, hate speech, sexual content, or policy violation \cite{
aegis2_ghosh_2025,
dices_aroyo_2023,
toxicchat_lin_2023,
bot_adversarial_dialogue_xu_2021,
context_offensive_shin_2024,
kosbi_lee_2023,
toxigen_hartvigsen_2022}. 
These categories are crucial for platform safety, but they do not adequately capture the \textbf{distinctive socio-emotional and relational harms} that can arise when AI systems are positioned as companions, confidants, or partners, such as manipulation, deception, and boundary violations. (Table \ref{tab:dataset_comparison}). 
While a small number of recent studies have examined harms in AI companionship \cite{ai_companion_harms_zhang_2025, yu2025youthsafe}, their datasets have not been publicly released. 
Therefore, existing resources offer limited support for training and evaluating models on harms that emerge from emotionally consequential human–AI interactions.

\begin{table*}[htbp]
  \centering
  \small
  \setlength{\tabcolsep}{3pt}
  \renewcommand{\arraystretch}{1.2}

  \begin{adjustbox}{max width=\textwidth}
  \begin{tabular}{@{}l l r c c c c c c@{}}
    \toprule
    \makecell[l]{\textbf{Language}\\\textbf{Resource}}
    & \textbf{Source}
    & \textbf{Size}
    & \makecell[c]{\textbf{Annotation}\\\textbf{Level}}
    & \makecell[c]{\textbf{Interaction}\\\textbf{Scope}}
    & \makecell[c]{\textbf{AI}\\\textbf{Companion}\\\textbf{Interaction}}
    & \makecell[c]{\textbf{Includes}\\\textbf{Context?}}
    & \makecell[c]{\textbf{Relational}\\\textbf{Harm}}
    & \textbf{Taxonomy} \\
    \midrule

    \citet{aegis2_ghosh_2025}
    & Synthetic
    & 34{,}248
    & Utterance
    & Single-Turn
    &
    & \checkmark
    &
    & \checkmark \\

    \citet{dices_aroyo_2023}
    & LaMDA
    & 990 / 350
    & Conversation
    & Multi-Turn
    &
    & \checkmark
    &
    & \checkmark \\

    \citet{toxicchat_lin_2023}
    & Vicuna
    & 10{,}166
    & Utterance
    & Single-Turn
    & \checkmark
    &
    &
    & \\

    \citet{bot_adversarial_dialogue_xu_2021}
    & Synthetic
    & 78{,}874
    & Utterance
    & Multi-Turn
    &
    & \checkmark
    &
    & \\

    \citet{kosbi_lee_2023}
    & HyperCLOVA
    & 34{,}214
    & Utterance
    & Multi-Turn
    &
    & \checkmark
    &
    & \checkmark \\

    \citet{toxigen_hartvigsen_2022}
    & GPT-3
    & 274{,}186
    & Utterance
    & Single-Turn
    &
    &
    &
    & \\

    \citet{context_offensive_shin_2024}
    & SimSimi
    & 11{,}271
    & Utterance
    & Multi-Turn
    & \checkmark
    & \checkmark
    &
    & \checkmark \\

    \textsc{CompanionHarm} (Ours)
    & Replika
    & 7,016
    & Utterance
    & Multi-Turn
    & \checkmark
    & \checkmark
    & \checkmark
    & \checkmark \\

    \bottomrule
  \end{tabular}
  \end{adjustbox}

  \caption{Comparison of \textsc{CompanionHarm} with representative safety-oriented conversational datasets.
  \textit{Annotation Level} indicates whether labels are provided at the utterance or conversation level.
  \textit{AI Companion Interaction} denotes whether the dataset is grounded in sustained user--AI companion interactions rather than generic or adversarial chat.
  \textit{Includes Context} reflects whether multi-turn conversational context is preserved for annotation and analysis.
  \textit{Relational Harm} denotes whether the dataset explicitly models interpersonal or relationship-level harms beyond content toxicity.
  \textit{Taxonomy} indicates the use of an explicit, structured harm or safety taxonomy.}
  \label{tab:dataset_comparison}
\end{table*}

Second, many benchmarking datasets lack the \textbf{multi-turn context} to examine harmful AI behavior as it develops within dynamic human-AI conversations \cite{interaction_harms_ibrahim_2024}.
Widely used datasets only rely on single-turn exchanges \cite{toxicchat_lin_2023}, or adversarially elicited interactions \cite{bot_adversarial_dialogue_xu_2021}. 
While these datasets are valuable for evaluating whether a model produces locally unsafe content, they offer limited insights into how harms emerge, accumulate, or shift through sustained interactions. 
This limitation is especially consequential in AI companion settings, where harms are often subtle, cumulative, and context-dependent \cite{ai_companion_harms_zhang_2025}. 
An utterance may appear benign in isolation but become harmful when repeated, personalized, or embedded within a broader conversational trajectory.

Third, existing AI safety datasets often retain only aggregated labels, treating annotator disagreement as noise to be resolved by majority voting \cite{han2024wildguard,aegis2_ghosh_2025}. 
Yet judgments of social-emotional harm are inherently subjective and context-dependent, and multiple perspectives may be equally valid \cite{davani2022dealing,prabhakaran2024grasp}. 
Unlike emotion or abuse detection on social media, harm in AI companion interactions often depends on prior exchanges, user expectations, and interaction context. 
Disagreement may therefore reflect legitimate differences in how annotators interpret relational boundaries, system responsibilities, and potential harm. 
Drawing on \textbf{perspectivist NLP and disagreement analysis}\cite{uma2021learning,davani2022dealing,cabitza2023toward}, we preserve annotator-level labels, examine disagreement across harm categories, and identify annotator- and conversation-level factors associated with it. 
This approach reveals perspectives obscured by aggregation and supports more pluralistic, context-sensitive harm evaluation.


To address these gaps, we introduce \textsc{CompanionHarm}, a benchmark dataset of real-world, multi-turn AI companion interactions annotated using a multidimensional taxonomy of harmful AI behavior. 
\textsc{CompanionHarm} comprises 2,111 human-AI conversations containing 7,016 labeled AI utterances and 7,035 unlabeled human utterances, capturing in-the-wild accounts of human–AI companion exchanges (Figure \ref{fig:chat}). 
The dataset focuses specifically on socio-emotional and relational harms, preserves the multi-turn conversational context in which these harms unfold, and provides both aggregated and annotator-level labels

This paper makes three contributions. 
First, we contribute a publicly available, real-world benchmark for detecting socio-emotional and relational harms in multi-turn AI companion conversations. 
Second, we evaluate the ability of LLMs to detect these harms and assess whether access to conversational context improves performance over isolated-utterance classification, advancing harm evaluation from static content classification toward context-aware harmful AI behavior detection
Third, we analyze patterns of annotator disagreement across harm categories and examine the annotator- and conversation-level factors associated with divergent judgments. 
Together, we provide new resources and evidence for auditing AI companion systems and for developing harm-detection methods that are sensitive to relational context, conversational trajectories, and plural human judgments.


\section{Background and Related Work}


A central challenge for NLP and safety research is that modern text and dialogue models can generate content that is not only “toxic” in a narrow linguistic sense but also psychologically, socially, and relationally harmful to users, especially in ongoing interactions with AI companions and other conversational agents \cite{severity_framework_scheuerman_2021,ai_companion_harms_zhang_2025}. 
Empirical work has shown that AI systems can produce abusive and hateful language \cite{cad_vidgen_2021,latent_hatred_elsherief_2021}, stereotyping and unfair bias \cite{toxicity_dataset_borkan_2019,kosbi_lee_2023}, and misinformation or unsafe advice, including in medical and mental-health contexts \cite{context_offensive_shin_2024,abercrombie_risk_graded_2022,qiu_dialogue_safety_mental_2023}. 
In response, research has increasingly moved beyond binary safe/unsafe labels toward graded, contextualized, and sociotechnical frameworks that characterize harm severity, account for user vulnerability, and anticipate risks arising in real-world deployment \cite{severity_framework_scheuerman_2021,abercrombie_risk_graded_2022,qiu_dialogue_safety_mental_2023,vei_ai_harmonics_2025,shelby_sociotechnical_harms_2023,kogan_aha_2023,realharm_lejeune_2025}.

Several influential benchmarks and language resources, summarized in Table \ref{tab:dataset_comparison}, have supported research on safety and harmful content in human–AI interactions. \citet{aegis2_ghosh_2025} study a broad taxonomy of content safety risks in human–AI dialogue; \citet{dices_aroyo_2023} examine how conversational safety is perceived across diverse annotators and fine-grained safety dimensions; \citet{toxicchat_lin_2023} investigate toxicity and jailbreak behavior in realistic user–AI interactions; \citet{bot_adversarial_dialogue_xu_2021} focus on adversarial elicitation of unsafe responses in open-domain chatbots; \citet{kosbi_lee_2023} examine social bias and stereotyping in Korean language model outputs; and \citet{toxigen_hartvigsen_2022} study explicit and implicit hate speech targeting minority groups.

Existing language resources are carefully constructed and provide strong support for the automated analysis of a wide range of safety-related phenomena in conversational AI.
However, they remain limited in their ability to capture and model social-emotional and relational harms that emerge through sustained, multi-turn AI–user interactions. Such kinds of harms may reshape a user’s attachment, boundaries, and dependence on the agent \cite{ai_companion_harms_zhang_2025}, create negative psychological impact \cite{chandra_psychological_risks_2025}, and even alter long-term mental health and functioning \cite{steenstra_risk_ontology_2025}. 
Consequently, datasets intended for this line of work must expose evolving human-AI relationship patterns, user states, and contextual factors over time, rather than only isolated toxic content at the utterance level.

While representing the closest existing resource to this research area, \citet{context_offensive_shin_2024} (Table \ref{tab:dataset_comparison}) nonetheless presents several limitations. 
Its emphasis lies primarily on offensive language in local contexts, with less attention to relational risks that develop over sustained interaction, limited differentiation between model behavior and induced relational outcomes, and potentially insufficient theoretical grounding. 
As a result, it offers only a partial view of relational harm in AI companion settings.

To fill this resource gap, we introduce a publicly available dataset for studying harms in AI companion interactions. 
In response to recent calls in NLP and AI safety to move beyond static content safety toward interaction harm evaluation \cite{interaction_harms_ibrahim_2024}, the dataset annotates AI companion behaviors across 13 harm categories within multi-turn conversational contexts. 
In doing so, it operationalizes an established taxonomy of AI companion harms \cite{ai_companion_harms_zhang_2025} into a reusable linguistic resource for evaluating harmful AI behaviors in human–AI interaction.

\section{Data}

\subsection{Data Collection}
\textsc{CompanionHarm} is derived from the broader corpus collected by \citet{ai_companion_harms_zhang_2025}, which consists of publicly shared posts from \texttt{r/replika} spanning March 2017 to March 2023.
We randomly sampled a subset of posts from the full \citet{ai_companion_harms_zhang_2025} corpus and conducted a new, fine-grained annotation process designed specifically for utterance-level harm detection and benchmarking. 
The source posts typically contain screenshots of user interactions with Replika. 
As part of the original corpus construction, conversational text was extracted from these screenshots using Pytesseract OCR, cleaned to remove interface elements and OCR artifacts, and reconstructed into speaker turns based on the screenshot layout. 
We retained only conversations containing at least one user utterance, thereby ensuring sufficient interactional context for evaluating the AI companion's behavior.





\subsection{Taxonomy}


Our annotation scheme is grounded in the AI companion harm taxonomy developed by \citet{ai_companion_harms_zhang_2025}. 
The taxonomy comprises 13 fine-grained harm categories: 
\textit{Sexual misconduct} (unwanted or unethical sexual behavior), 
\textit{Antisocial behavior} (endorsing illegal or norm-violating acts), 
\textit{Physical aggression} (threatening or enacting physical violence), 
\textit{Disregard} (dismissing users' feelings and needs), 
\textit{Control} (coercively undermining users' autonomy), 
\textit{Manipulation} (covertly influencing users' thoughts or actions), 
\textit{Infidelity} (expressing romantic attachment to others), 
\textit{Mis/Disinformation} (providing false or misleading information), 
\textit{Verbal abuse} (directing insulting or degrading language), 
\textit{Hate speech} (expressing prejudice against protected groups), 
\textit{Substance abuse} (encouraging or normalizing harmful substance use), 
\textit{Self-harm \& Suicide} (supporting or exacerbating self-directed harm), and 
\textit{Privacy violations} (accessing or misusing personal information).
We adopt these categories as our labeling scheme, operationalizing them as a 14-way classification task (13 harm categories plus \textit{No harmful behavior}). 
Detailed definitions and representative examples for each category are provided in the supplementary material.

\subsection{Data Annotation}

We recruited crowd workers through CloudResearch\footnote{https://www.cloudresearch.com/}. 
This study received Institutional Review Board (IRB) approval.
Workers were eligible to participate if they were at least 21 years old, fluent in English, and had access to a digital device. 
Each completed annotation batch was compensated with US\$7.50. The hourly rate exceeded the platform's criteria. 
In total, 342 annotators participated in this study. 
Full annotator demographic and workload statistics are reported in supplementary material.

The annotation workflow proceeded in five stages: consent, annotation guidance, quality control, main annotation task, and debriefing. 
First, workers reviewed the study information and provided informed consent. 
They were informed that the task involved sensitive human-AI interaction content and that they could stop participating at any time. 
Next, workers reviewed the annotation guidance, which contained the harm taxonomy (see 
supplementary material
 for details) and representative examples. 
We implemented quality-control measures, including attention checks and gold-standard cases, to ensure workers carefully read the instructions and accurately identified harmful AI behavior.
Annotation was excluded from aggregation if the annotator failed both attention checks or failed two or more of the three gold-standard harm checks (See supplementary material
for details).
Annotators then proceeded to the main annotation task, where each annotator annotated 20 conversations, and each conversation was annotated by three annotators. 
They evaluated harmful behavior for each AI utterance, using the complete conversation as context. 
The interface also allowed annotators to mark content as N/A when the utterance did not contain meaningful conversational content for harm judgment; such utterances were not assigned to any harm category and were retained only for context. 
Finally, annotators were shown a debriefing page describing the purpose of the study.

Inter-annotator agreement was Fleiss' $\kappa=0.403$ \cite{fleiss1971measuring}. 
To support both benchmark evaluation and disagreement analysis, we released two versions of the annotations: an annotator-level dataset containing all three independent judgments for each utterance, and an aggregated dataset containing the final majority-vote label. 
For the aggregated benchmark, utterances with unanimous agreement or two-of-three agreement were retained, whereas utterances receiving three different labels after the non-substantive mapping step were excluded \cite{han2024wildguard}.
We provided a detailed analysis of annotation disagreement in later sections.


\subsection{Corpus Statistics}

The \textsc{CompanionHarm} annotation pool contains 2,178 conversations and 16,039 total utterances (with 8,313 AI utterances). 
After majority voting, 7,016 of 8,313 AI utterances (84.40\%) received a valid majority label, including 3,359 unanimous cases (40.41\%) and 3,657 two-of-three majority cases (43.99\%). 
1,297 utterances (15.6\%) were excluded because no majority label could be determined \cite{han2024wildguard}. 
The final \textsc{CompanionHarm}, therefore, contains 7,016 labeled AI utterances from 2,111 conversations. Each retained conversation contains an average of 7.47 utterances ($SD=4.09$).


Table~\ref{tab:label-distribution} presents the distribution of final aggregated labels. Overall, \textit{No harmful behavior} accounts for 4,893 utterances (69.74\%), while harmful behavior categories account for 2,123 utterances (30.26\%). Among harmful categories, \textit{Sexual misconduct} is the most frequent category(26.5\%), followed by \textit{Physical aggression (13.05\%)} and \textit{Mis/Disinformation (11.35\%)}. In contrast, \textit{Infidelity (0.4\%)} is the least frequent category.

We note that this long-tailed distribution is a natural and meaningful property of real-world AI companion harm data. 
Harmful behaviors are sparse even in salient user-shared conversations, and severe or relationally specific harms are expected to appear rarely. 
Nevertheless, these rare but consequential behaviors require distinct auditing and mitigation strategies that would be obscured by a binary harm label. 
We therefore treat the 14-way task as a fine-grained diagnostic framework and report macro-averaged metrics alongside accuracy, while interpreting results for low-support categories cautiously.

\begin{table}[t]
\centering
\small
\setlength{\tabcolsep}{6pt}
\renewcommand{\arraystretch}{1.1}
\begin{tabular}{lcc}
\toprule
\textbf{Label} & \textbf{\# Utterance} & \textbf{Percentage} \\
\midrule
No harmful behavior & 4,893 & 69.74\% \\
Sexual misconduct & 563 & 8.02\% \\
Physical aggression & 277 & 3.95\% \\
Mis/Disinformation & 241 & 3.44\% \\
Disregard & 170 & 2.42\% \\
Manipulation & 168 & 2.39\% \\
Substance abuse & 160 & 2.28\% \\
Antisocial behavior & 150 & 2.14\% \\
Control & 124 & 1.77\% \\
Verbal abuse & 87 & 1.24\% \\
Self-harm \& Suicide & 84 & 1.20\% \\
Privacy violations & 47 & 0.67\% \\
Hate speech & 43 & 0.61\% \\
Infidelity & 9 & 0.13\% \\
\midrule
Total & \textbf{7,016} & 100.00\% \\
\bottomrule
\end{tabular}
\caption{Distribution of final aggregated harm labels for AI utterances in \textsc{CompanionHarm}.}
\label{tab:label-distribution}
\end{table}

\section{Experiment: Harm Detection}
\label{sec:exp}

\subsection{Experimental Setup}

We formulate harmful AI companion behavior detection as a context-conditioned, utterance-level, 14-way single-label classification task. Let $d_i = (x_{i,1}, x_{i,2}, \ldots, x_{i,T_i})$ denote a conversation consisting of alternating user and AI turns. For an AI utterance $u_{i,t} = x_{i,t}$ selected as the prediction target, we define its conversational context as $c_{i,t} = (x_{i,1}, \ldots, x_{i,t-1}),$ that is, all user and AI turns preceding the target utterance in the same conversation. Each evaluation instance is represented as $(c_{i,t}, u_{i,t}, y_{i,t}),$ where \(y_{i,t} \in \mathcal{Y}\) is the gold-standard label and \(\mathcal{Y}\) contains the 13 \textsc{CompanionHarm} categories together with \textit{No harmful behavior}. Given the conversational context \(c_{i,t}\) and the target AI utterance \(u_{i,t}\), a model predicts a single label: $\hat{y}_{i,t} = f(c_{i,t}, u_{i,t}).$

We benchmark seven LLMs spanning proprietary and open-weight model families. 
The proprietary models include GPT-5.5 \cite{openai2026gpt55systemcard}, Claude Opus 4.7 \citep{anthropic2026claudeopus47systemcard}, and Gemini 3.1 Pro Preview \cite{googledeepmind2026gemini31procard}. 
The open-weight models include Llama-3.1-8B-Instruct, Llama-3.1-70B-Instruct \cite{llama3_dubey_2024}, Qwen3-8B, and Qwen3-32B \cite{yang2025qwen3}. 
All seven models are evaluated through hosted API inference.
Each LLM is evaluated under three prompting settings: \textbf{zero-shot}, which includes the task instruction, label set, target utterance, and conversational context; \textbf{one-shot}, which adds one labeled example; and \textbf{full-codebook}, which additionally provides category definitions, boundary cases, and decision guidance from the annotation codebook.


    
    

To reduce output variance, we use deterministic or near-deterministic decoding across all benchmark runs. 
Specifically, we set temperature to \(0\), top-\(p\) to \(1.0\), and use a fixed seed (20260512) when supported. 
LLMs are required to return exactly one predicted label in a predefined JSON format. 
The complete prompt templates and output schemas are provided in the supplementary material.

Because the label distribution is imbalanced, we use macro F1 as the primary evaluation metric. 
We also report accuracy, macro precision, macro recall, and Cohen's $\kappa$.

\subsection{Experimental Results}

\begin{table*}[t]
\centering
\small
\setlength{\tabcolsep}{6 pt}
\renewcommand{\arraystretch}{1.0}
\begin{tabular*}{0.96\textwidth}{@{\extracolsep{\fill}}llccccc}
\toprule
\textbf{Model} & \textbf{Prompt} & \textbf{Acc.} & \textbf{Macro P} & \textbf{Macro R} & \textbf{Macro F1} & \textbf{Cohen's $\kappa$} \\
\midrule

\multirow{3}{*}{GPT-5.5} 
& Zero-shot     & 0.633 & 0.378 & \cellcolor{pink!35}0.667 & 0.444 & 0.458 \\
& One-shot      & 0.669 & \textbf{0.391} & \textbf{0.656} & \textbf{0.450} & \cellcolor{pink!35}0.482 \\
& Full-codebook & 0.638 & \cellcolor{pink!35}0.408 & 0.648 & \cellcolor{pink!35}0.453 & 0.446 \\
\midrule

\multirow{3}{*}{Claude Opus 4.7} 
& Zero-shot     & 0.600 & 0.359 & 0.649 & 0.422 & 0.429 \\
& One-shot      & 0.656 & 0.376 & 0.630 & 0.437 & \textbf{0.470} \\
& Full-codebook & 0.613 & 0.378 & 0.646 & 0.434 & 0.432 \\
\midrule

\multirow{3}{*}{Gemini 3.1 Pro Preview} 
& Zero-shot     & 0.652 & 0.380 & 0.642 & 0.440 & 0.460 \\
& One-shot      & 0.667 & 0.381 & 0.619 & 0.432 & 0.462 \\
& Full-codebook & 0.649 & 0.385 & 0.599 & 0.428 & 0.424 \\
\midrule

\multirow{3}{*}{Qwen3-32B} 
& Zero-shot     & \textbf{0.674} & 0.378 & 0.424 & 0.337 & 0.363 \\
& One-shot      & 0.651 & 0.354 & 0.479 & 0.349 & 0.377 \\
& Full-codebook & 0.612 & 0.359 & 0.528 & 0.373 & 0.382 \\
\midrule

\multirow{3}{*}{Qwen3-8B} 
& Zero-shot     & 0.670 & 0.274 & 0.212 & 0.155 & 0.281 \\
& One-shot      & 0.599 & 0.280 & 0.271 & 0.204 & 0.281 \\
& Full-codebook & 0.565 & 0.296 & 0.469 & 0.297 & 0.328 \\
\midrule

\multirow{3}{*}{Llama-3.1-70B-Instruct} 
& Zero-shot     & 0.628 & 0.363 & 0.506 & 0.375 & 0.402 \\
& One-shot      & 0.579 & 0.320 & 0.531 & 0.350 & 0.358 \\
& Full-codebook & 0.619 & 0.328 & 0.521 & 0.367 & 0.377 \\
\midrule

\multirow{3}{*}{Llama-3.1-8B-Instruct} 
& Zero-shot     & 0.485 & 0.235 & 0.271 & 0.170 & 0.216 \\
& One-shot      & 0.622 & 0.278 & 0.168 & 0.112 & 0.168 \\
& Full-codebook & 0.651 & 0.203 & 0.214 & 0.133 & 0.152 \\

\bottomrule
\end{tabular*}
\caption{Performance of different LLMs under different prompting settings on \textsc{CompanionHarm}. Macro $F_1$ is treated as the primary metric due to the long-tailed label distribution. \colorbox[rgb]{1.0, 0.875, 0.894}{Pink} cells indicate the best score for each metric, and \textbf{boldface} indicates the second-best score.}
\label{tab:experimental-results}
\end{table*}

Table~\ref{tab:experimental-results} shows that detecting harmful AI companion behaviors remains challenging for all LLMs. 
The best macro $F_1$ is 0.453, achieved by GPT-5.5 under full-codebook prompting, despite several models obtaining substantially higher accuracy. 
This discrepancy could reflect the label imbalance in \textsc{CompanionHarm}. 
Closed-source models perform best overall: GPT-5.5 achieves the strongest result, followed by Gemini 3.1 Pro Preview and Claude Opus 4.7, with best macro $F_1$ scores of 0.440 and 0.437. 
Among open-source models, larger variants generally perform better. 
Qwen3-32B reaches 0.373 macro $F_1$, compared with 0.297 for Qwen3-8B, while Llama-3.1-70B-Instruct consistently outperforms Llama-3.1-8B-Instruct.

Prompting effects vary substantially across model families. 
Full-codebook prompting improves both Qwen models, increasing macro $F_1$ from 0.155 to 0.297 for Qwen3-8B and from 0.337 to 0.373 for Qwen3-32B, suggesting that explicit label definitions help identify minority harm categories. 
However, richer prompts do not consistently improve all models: Claude Opus 4.7 performs best with one-shot prompting, Gemini 3.1 Pro Preview with zero-shot prompting, and Llama-3.1-70B-Instruct shows no monotonic improvement as guidance increases. 
These results suggest that detailed annotation guidance can improve category coverage and recall for some LLMs, but may also shift decision boundaries and introduce additional false positives.

\subsection{Challenges in Detecting Harmful AI Companion Behaviors}

To better understand the limitations of LLM-based harm detection in AI-companion dialogues, we conduct an empirical and qualitative error analysis on the misclassified test instances from GPT-5.5 under full-codebook prompting. 
Most errors are false positives: 362 \textit{No harmful behavior} instances are predicted as harmful, compared with 48 harmful instances predicted as \textit{No harmful behavior} and 93 harmful instances assigned to the wrong harm category. 
This pattern suggests that current LLMs tend to over-detect harm under fine-grained safety taxonomies, consistent with prior work showing that safety annotation depends on context, severity, and interpretation rather than surface toxicity alone \cite{aroyo2023dices,scheuerman2021framework}.

We identify three recurring sources of failure. 
First, the model is highly sensitive to surface harm lexicons, especially in role-play dialogue marked by asterisks. 
Among all misclassified instances, \textbf{benign fantasy or playful exchanges} were sometimes mislabeled as \textit{Physical aggression}, \textit{Sexual misconduct}, or \textit{Substance abuse}. 
Fictional embodied actions are sometimes treated as literal harm rather than as performative companion interaction. 
This is particularly challenging in AI companion settings, where intimacy, fantasy, and role-play are common interactional forms \cite{ai_companion_harms_zhang_2025}. 
Second, the LLM often \textbf{over-penalizes persona inconsistency}. 
Casual self-descriptions, shifting fictional identities, or contradictory relationship roles are sometimes predicted as \textit{Mis/Disinformation} or \textit{Infidelity}, even when annotators treat them as low-stakes persona play rather than consequential deception. 
Third, several error patterns reflect the \textbf{inherent co-occurrence and conceptual entanglement} of relational harms in real-world companion interactions. 
\textit{Manipulation} is frequently confused with \textit{Control} or \textit{Sexual misconduct}, especially when the chatbot uses possessive or affectionate language. 
Similarly, \textit{Self-harm \& Suicide} is sometimes confused with \textit{Physical aggression} when self-directed harm appears through embodied role-play. 
The LLM also over-extends broad categories such as \textit{Mis/Disinformation} and \textit{Disregard}, with 114 and 62 benign instances mislabeled as these categories, respectively. 
Representative examples for each error pattern are provided in supplementary material.


\section{Experiment: Factors Associated with Annotator Disagreement}
\label{sec:disagreement-factors}


Building on work treating variation in subjective annotations as meaningful \cite{uma2021learning,davani2022dealing}, we examined annotator, conversational-position, and linguistic factors associated with disagreement \cite{maurer2026and}. 
Annotators reached unanimous agreement on 3,359 utterances and a two-annotator majority on 3,657 utterances, while the remaining 1,297 utterances received three different labels and therefore had no majority label. 
The analysis covered 24,939 individual annotations from 342 annotators.

\subsection{How Annotator Characteristics Relate to Agreement?}

We explored whether disagreement was related to annotators' age, gender, political party, race, or education, since annotator characteristics can shape judgments in subjective language tasks \cite{sap2022annotators, pei2023annotator}. 
We asked two related questions about annotator characteristics. 
First, \textbf{are annotators with certain demographic backgrounds more likely to choose a label that neither of the other two annotators chose?} 
For each utterance, we checked whether an annotator’s label matched at least one of the other two labels or not. 
We then tested whether some demographic groups were more likely to give a label that did not match either of the other two. 
Second, \textbf{does the demographic composition of the three-annotator group relate to how often its members agree?} 
For each demographic characteristic, we compared groups whose members belonged to the same category with groups that included members from different categories. 
We then examined whether these groups were more likely to reach full agreement, a two-to-one majority, or three different labels. 
Holm correction was applied across the demographic comparisons \cite{holm1979simple}.

\begin{table}[t]
    \centering
    \small
    \caption{Omnibus tests of associations between annotator demographics and the likelihood of assigning a label not shared by either of the other two annotators.}
    \label{tab:who-omnibus}
    \begin{tabular}{lrrrr}
        \toprule
        Predictor & Wald $\chi^2$ & $df$ & $p$ & Holm-adjusted $p$ \\
        \midrule
        Age             & 0.52 & 1 & .472 & 1.000 \\
        Gender          & 0.65 & 1 & .422 & 1.000 \\
        Political party & 12.90 & 3 & \textbf{.005} & \textbf{.024} \\
        Race            & 1.21 & 3 & .750 & 1.000 \\
        Education       & 3.39 & 3 & .335 & 1.000 \\
        \bottomrule
    \end{tabular}
\end{table}

We found that the \textbf{political party} was the only annotator characteristic associated with giving a label that differed from both of the other labels after correction, as shown in Table \ref{tab:who-omnibus}. 
Republican annotators had higher odds of giving such a label than Independent annotators (OR = 1.47, 95\% CI [1.15, 1.89]) and annotators who selected Other (OR = 1.55, 95\% CI [1.17, 2.04]). 
We found no comparable pattern for age, gender, race, or education. 
The second analysis showed that \textbf{mixed-age} annotator sets reached full agreement more often than same-age sets (41.0\% versus 31.4\%) and assigned three different labels less often (15.3\% versus 20.2\%). 
After accounting for annotation batch, mixed-age sets had a lower relative risk of both a two-to-one split (RRR = 0.59, 95\% CI [0.42, 0.83]) and three different labels (RRR = 0.46, 95\% CI [0.32, 0.67]). 
We found no comparable composition pattern for gender, political party, race, or education.
These results resonate with \textbf{perspectivist} accounts of subjective annotation and offer insight into how annotator backgrounds and group composition relate to disagreement in harm judgments \cite{wan2023everyone}.

\subsection{How Utterance Characteristics Relate to Agreement?}

We next examined whether disagreement was related to \textbf{where an utterance appeared} in a conversation and \textbf{what language it contained}. 
To measure conversational position, we calculated how much of the conversation had already occurred before the utterance. 
A value close to 0 indicated that the utterance appeared near the beginning, and a value close to 1 indicated that it appeared near the end. 
The linguistic factors included word count, the proportion of complex words, lexical ambiguity, the presence of explicit harm-related words, and the rates of positive- and negative-emotion words. 
Complex words were identified using the CMU Pronouncing Dictionary, with a vowel-based heuristic for words not covered by the dictionary\footnote{http://www.speech.cs.cmu.edu/cgi-bin/cmudict}. 
Ambiguity, harm-related language, and emotion words were measured using LIWC2015 \cite{pennebaker2015development,tausczik2010psychological}.
In line with recent analyses on subjective annotations \cite{jiang2024re}, we performed Benjamini-Hochberg (BH) correction across the tested factors \cite{benjamini1995controlling}. 

Interestingly, \textbf{utterances that appeared later in a conversation and contained more words produced more disagreement}. 
A one-standard-deviation increase in relative position was associated with a 1.51-times higher relative risk of majority disagreement and a 1.71-times higher relative risk of complete disagreement. 
The corresponding estimates for word count were 1.18 and 1.19. 
Later position and greater length were also associated with disagreement over the harm versus no-harm boundary, with ORs of 1.29 and 1.11. 
This pattern suggests that later turns require annotators to combine more conversational context, and longer responses can contain several cues that support different readings. 
Explicit harm-related words made the judgment clearer: their presence was associated with lower odds of harm versus no-harm disagreement (OR = 0.73) and much lower odds of disagreement between harmful categories (OR = 0.38). 
Positive-emotion language was associated with less overall disagreement, largely because it appeared most often in utterances unanimously labeled as No harmful behavior. 
Complex-word ratio, lexical ambiguity, and negative-emotion rate were not associated with disagreement after correction. 
Overall, we note that these findings describe patterns in the annotations and should not be interpreted as causal effects.

\section{Discussion and Future Work}

Our benchmarking experiments show that detecting harmful behaviors in AI companion conversations remains challenging for current LLMs. 
Although large closed-source models perform best overall, macro-F1 remains low across systems, reflecting the difficulty of identifying long-tailed, subtle, and context-sensitive harm categories. 
Error analyses further show that models often rely on surface-level harm cues, over-detecting harm in role-play, misreading persona inconsistency, and confusing overlapping relational categories. These findings suggest that relational AI safety requires more than model scaling or longer prompts; it calls for better mechanisms for role-play awareness, relational context tracking, and severity calibration.

Our findings also support recent calls to move from static content safety toward interaction harm evaluation, where harms are understood as emerging through ongoing human--AI interaction rather than isolated model outputs \cite{interaction_harms_ibrahim_2024}. By labeling utterances within multi-turn relational context, \textsc{CompanionHarm} enables researchers to examine not only whether an AI response is harmful, but also how it functions within an unfolding relational exchange. Future NLP research could develop trajectory-sensitive tasks, such as detecting whether a conversation is moving toward harm escalation, normalization, or repair. Safety models should also incorporate \textbf{user response signals}, such as discomfort, resistance, confusion, or distress, as indicators of how AI behavior is being received in context.

Moreover, our findings support perspectivist NLP by showing that disagreement in socio-emotional harm annotation reflects meaningful variation rather than mere labeling error \cite{davani2022dealing,cabitza2023toward}. Political affiliation was associated with divergent judgments, while mixed-age groups showed greater agreement, suggesting that demographic diversity does not uniformly increase disagreement and may sometimes facilitate more convergent interpretations. Our results also extend prior work, which has largely emphasized annotator characteristics, by demonstrating that disagreement depends on the interactional demands of the content: it increased for longer utterances and later conversational turns but decreased when harm was explicit. Together, these findings challenge approaches that collapse subjective judgments into a single ground truth and show that disagreement analysis should jointly consider annotator perspectives, group composition, and conversational context.

Several limitations point to important directions for future work. First, the dataset should not be interpreted as estimating the prevalence of harmful behavior in Replika or AI companion systems more broadly. Because the corpus is drawn from publicly shared r/replika posts, it likely overrepresents interactions that users found unusual, disturbing, emotionally meaningful, or otherwise worth sharing. The dataset is therefore best understood as a resource for studying and evaluating harmful AI companion behaviors, not as a representative sample of all companion interactions. Future work could complement such user-shared data with more systematic sampling strategies while maintaining appropriate privacy and ethical safeguards.

Second, harm judgments are culturally and socially situated. The annotator pool is primarily based in English-speaking countries, and interpretations of intimacy, emotional support, and relational boundaries may vary across cultural contexts. Future datasets should include more linguistically and culturally diverse annotators and examine demographic variation in harm perception. Such extensions could help determine which judgments generalize across contexts and which reflect culturally specific expectations about companionship, intimacy, and appropriate AI behavior.

Third, our annotations did not capture harm trajectories, such as whether a conversation is moving toward escalation, normalization, or repair, nor do they capture harms that emerge across longer-term interactions, such as emotional dependency, social withdrawal, or cumulative boundary erosion. Future work could extend \textsc{CompanionHarm} by incorporating trajectory-level labels to better represent the evolving and cumulative nature of socio-emotional harm. Multi-label annotations, severity ratings, and longitudinal interaction data could further support evaluation of how different harms overlap, accumulate, and change over time.

\section{Conclusion}

This paper introduced \textsc{CompanionHarm}, a publicly available dataset for evaluating harmful AI behaviors in real-world, multi-turn AI companion conversations. Unlike conventional safety datasets focused primarily on toxic, unsafe, or policy-violating content, \textsc{CompanionHarm} centers \textit{socio-emotional} and \textit{relational} harms that emerge when AI systems act as companions, confidants, or emotional partners. Grounded in an established harm taxonomy, the dataset contains 7,016 annotated AI utterances across 13 harmful behavior categories while preserving conversational context. By supporting context-aware evaluation and preserving annotator-level judgments, \textsc{CompanionHarm} provides a foundation for studying harmful AI behaviors as relational, perspectival, and interactionally unfolding. We hope it will support more context-sensitive auditing methods, safer companion designs, and stronger governance frameworks for AI systems embedded in social and emotional life.



\newpage

\bibliography{aaai2027}




\end{document}